**Cross-Axis Weighted Harmonic Method: A Frequency-Domain Approach for Enhanced Resolution in Magnetic Particle Imaging.**

Abuobaida M.khair[a,b], Wenjing Jiang[a], Moritz Wildgruber[c], Wenjun Xia[d,*], Xiaopeng Ma[a,]

1. Introduction

Magnetic Particle Imaging (MPI) is an innovative and rapidly evolving medical imaging technique that provides a unique means of visualizing the interior of the body. MPI directly detects special nanoparticles called magnetic nanoparticles (MNPs). These tiny tracers can be injected into the body and accurately tracked and measured, allowing MPI to pinpoint their location and concentration in real time[1][2].

MPI leverages the unique behavior of MNPs when they are exposed to specific magnetic fields, i.e., the selection field (static) and the drive field (oscillating), created by a carefully designed arrangement of coils inside the scanner. When the magnetic field interacts with these nanoparticles, they produce a distinct response that is captured by receiving coils. This signal forms the foundation for creating detailed images, showing where the tracers are inside the body[3]. When magnetic fields are applied to MNPs, the tiny magnetic moments inside each particle try to align with the field, creating a net magnetization. This change generates an induced voltage signal that MPI scanners can detect and utilize to create detailed images of the particle locations. However, the response of the particles is not instant—they take a brief, predictable moment to react, known as relaxation time. During this time, the particles can respond in two ways: they might physically rotate (called Brownian relaxation) or their internal magnetic direction might shift (known as Néel relaxation). Often, especially in fluid environments, both types of movement happen together, and how they respond depends on how quickly the magnetic field changes. To describe this behavior, the Langevin theory is used, which assumes the particles are always in balance with their surroundings [4][5][6].

To reconstruct an image in MPI, we must figure out where and how many tracer particles are present. This is done by solving equations that link the particle concentration to the signals picked up by the scanner's receive coils[7]. The system matrix is the key to this process—it acts like a map, showing how the particles' locations relate to the signals they produce. This matrix is usually analyzed in the frequency domain and depends on the scanning method and how the particles respond to changing magnetic fields. While there are theoretical models that describe how these particles behave, accurately simulating their response in practice is still challenging. That is why the most reliable way to get this system matrix is through a careful and time-consuming calibration process using real measurements[8][9].

The strength of the induced voltage signal detected is closely tied to how quickly the magnetization of the MNPs changes. This is because the induced voltage is proportional to the rate of change of the magnetic flux, which itself depends on how fast the MNPs' magnetization responds—essentially, how rapidly it oscillates with the drive field. In simple terms, higher drive frequencies lead to a faster change in magnetization, resulting in stronger signals and better signal-to-noise ratio (SNR)[10].

However, drive frequency does not just affect signal strength—it also influences spatial resolution. This is because of the point spread function (PSF), which describes how a single point of MNPs appears in the final image. Higher drive frequencies improve SNR and speed up imaging. Still, they can also cause image blurring as the magnetization response becomes more spread out, causing the PSF to be widened, which results in a larger full-width at half-maximum (FWHM), and also because MNPs cannot respond instantly due to the relaxations. If the frequency is too high, the particles cannot keep up, and the image becomes less sharp[11][5].

Contrarily, using a lower drive frequency gives the particles more time to respond accurately, which can reduce blurring and improve spatial resolution—but it also slows down the imaging process and weakens the signal[12][13].

While the selection field, which is responsible for creating a field-free point (FFP), plays an important role in spatial localization by saturating MNPs outside the region of interest, our focus here is on the drive field on the induced voltage signal.

In this study, we aim to bridge the gap between realistic MPI signal modeling and efficient image reconstruction by building a full simulation pipeline that closely mirrors the physical processes in an MPI scanner. The key idea is to isolate and use the most informative parts of the induced voltage signal—specifically, higher-order harmonics that carry rich spatial information—through a method we call Cross-Axis Harmonic Analysis (CAHA). By carefully analyzing the induced voltage signals from all three spatial directions, CAHA helps to extract only the most meaningful components, leading to higher-resolution images[14][15].

Our framework simulates realistic MNPs behavior and system responses and explores how different drive field frequencies impact signal quality and spatial resolution. This approach allows us to understand and optimize the trade-offs in MPI and to evaluate the benefits of harmonic-focused signal processing techniques.

The main contributions to this work are:

- An end-to-end MPI simulation framework that includes field generation, MNPs modeling, signal acquisition, system function computation, system matrix construction, and image reconstruction using a 3D vascular phantom.
- Introduction of Cross-Axis Harmonic Analysis (CAHA), a directional signal processing method that extracts key harmonic components from the induced voltage signals along the x, y, and z directions, improving signal quality and image fidelity.
- Unlike the prior works that vary drive field amplitude, we analyze the impact of drive field frequency on the induced voltage signal strength and resolution using CAHA to optimize the trade-off between image clarity and the signal strength.
- Evaluation of the reconstructed image quality using CAHA-processed signals compared to reconstructions using the original induced signals across various drive frequencies.

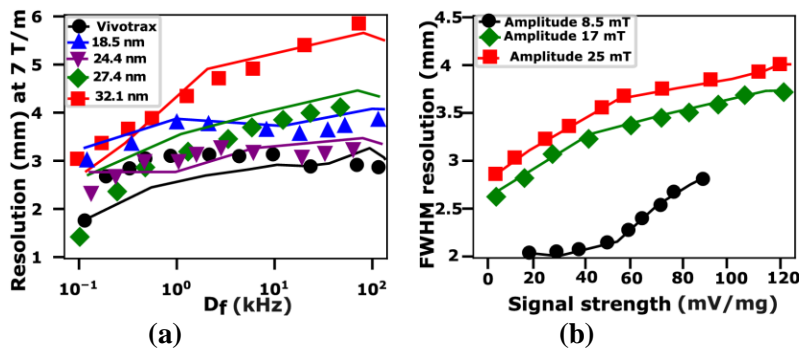

(a)       (b)

**Fig.1.** Experimental results reported in [16][17][18] (a) demonstrate that varying the drive frequency ($D_f$) between 0.5 kHz and 100 kHz at different core sizes of MNPs, while keeping the drive field amplitude constant at 20 mT/$\mu_0$, significantly influences system performance. (b) The signal strength and the spatial resolution relation for different applied field strengths.

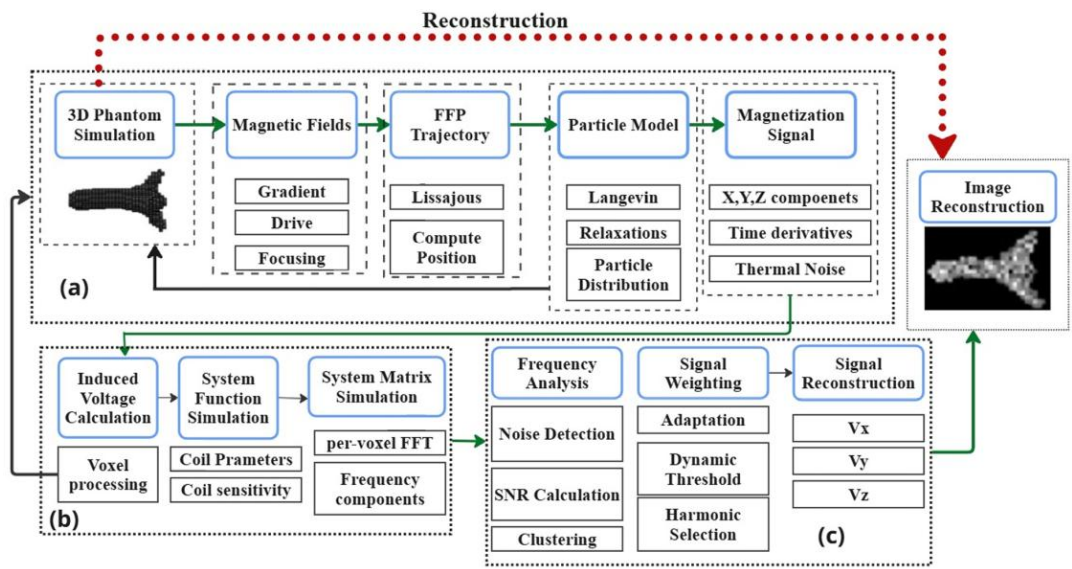

**Fig. 2.** Overview of the simulation workflow. (a) Illustrates the core simulation steps, starting with creating a numerical phantom and generating magnetic fields. (b) Shows how the induced voltage signals are simulated along with the system function. (c) Describes the signal processing pipeline using CAHA to extract meaningful information for image reconstruction. The green arrows indicate the sequence of the simulation model, and the black arrows indicate that the block is related to another block.

**Table1.** Comparison of key methodologies in MPI simulation and signal processing from selected literature. Checkmarks (✔) indicate features addressed in each study, while crosses (✘) denote absent elements. The table highlights gaps in cross-axis analysis and noise-adaptive filtering across existing works.

| Paper | 3D MPI Simulation | Signal-Domain Harmonic Filtering | Cross-Axis Analysis | Drive Field Analysis | Spatial Resolution Enhancement | Noise-Adaptive Filtering |
|---|---|---|---|---|---|---|
| Optimization of Drive Parameters[16] | ✔ (via relaxometer) | ✔ (harmonic shift tracked) | ✘ | ✔ | ✔ | ✘ |
| A 3D MPI Simulation [19] | ✔ | ✘ | ✘ | ✘ | ✔ | ✘ |
| Weighted harmonic[20] | ✘ | ✔ | ✘ | ✘ | ✔ | ✘ |
| Modular Simulation framework[15] | ✔ | ✘ | ✘ | ✘ | ✔ | ✘ |
| Perpendicular Signal Transform[21] | ✘ | ✔ (3D harmonic) | ✔ (perpendicular only) | ✘ | ✔ | ✘ |
| SNR-Based frequency selection[22] | ✔ (3D Simulation + Experimental) | ✔ (SNR-peak frequency select) | ✘ | ✘ | ✔ | ✔ (SNR aware select) |
| Mixing Excitation spatial encoding[23] | ✓ (2D+in vivo) | ✓ (sideband harmonics) | ✓ (multi-channel) | ✗ | ✓ | ✗ |
| **ours** | ✓ | ✓ | ✓ (x, y, z axes) | ✓ | ✓ (via SNR-rich harmonics) | ✓ (Adaptive SNR +) |

## 2. Methods

We developed a comprehensive simulation framework for an end-to-end MPI system to reconstruct a high-resolution image of a 3D numerically simulated phantom. The core idea is to extract and reconstruct meaningful signal components using Cross-Axis Harmonic Analysis (CAHA), resulting in improved image fidelity. The simulation model is divided into three major stages. First, we conduct the basic simulation, which includes simulating the magnetic drive and selection fields and creating a numerical phantom that depicts the spatial distribution of magnetic nanoparticles. The second stage uses the phantom's magnetization response to compute the induced voltage signals and accompanying system functions, capturing the spatial encoding capabilities of the simulated MPI system. In the final stage, we model the system matrix and use CAHA on the induced voltage signals to separate key harmonic components. These processed signals are then utilized to reconstruct a high-resolution image of the phantom by comparing it to the original signal without employing the CAHA approach at various drive frequencies. This structured approach enables us to realistically replicate the MPI signal pathway and assess the efficacy of harmonic-based signal processing for image reconstruction, and the CAHA approach leverages the orthogonality of the fields to reconstruct multi-planar views.

### 2.1. Simulation framework

The initial step focuses on modeling the physical environment and the interaction of magnetic fields with the MNPs. This stage involves creating a synthetic vascular structure phantom, configuring magnetic fields, and simulating MNPs' behavior in response to the applied fields. A synthetic 3D vascular phantom is created to resemble the geometry of branching blood vessels (see Fig.3). This phantom represents the spatial distribution of magnetic nanoparticles and functions as a test bed for simulating field interactions. The phantom is made up of many voxels and interconnected branches, resulting in a realistic vascular topology suited for MPI signal modeling and image reconstruction. This phantom is considered in all simulation steps to ensure the consistency of calculations.

Three basic components: a static selection field, a time-varying drive field, and a focusing field. The selection field, $H_s(r)$ is defined using a gradient field centered at the origin to generate a field-free point (FFP). It is mathematically expressed as:

$$H_s(r) = G_x x \hat{\imath} + G_y y \hat{\jmath} + G_z z \hat{k} \qquad (1)$$

where $r = (x, y, z)$ denotes the spatial coordinates, and $G_x, G_y, G_z$ are the gradient strengths along each axis. The drive field, $D_f(t)$, oscillates sinusoidally to move the FFP over time, defined as:

$$D_f(t) = A_x \sin(2\pi f_x t)\hat{\imath} + A_y \sin(2\pi f_y t)\hat{\jmath} + A_z \sin(2\pi f_z t)\hat{k} \qquad (2)$$

where $A_x, A_y, A_z$ and $f_x, f_y, f_z$ are the amplitudes and frequencies of the drive fields along x, y and z axis, respectively, and $\hat{\imath}, \hat{\jmath}, \hat{k}$ are the unit vectors in a 3D Cartesian coordinate x, y, and z, respectively. The resulting trajectory of FFP traces a 3D Lissajous curve, which is used to scan the imaging volume. A typical focusing field is used to slowly shift the FFP to cover the FOV modeled as:

$$H_F(t) = B_x(t)\hat{i} + B_y(t)\hat{j} + GB_z(t)\hat{k} \qquad (3)$$

where $B_x(t), B_y(t), B_z(t)$ are slow-varying functions that define how the FFP shifted over time.

Thus, the total magnetic field experienced at any spatial location within the phantom is computed as:

$$H(r,t) = H_s(r) + D_f(t) + H_F(t) \qquad (4)$$

The MNPs' magnetic response $M(r,t)$ is defined using the Langevin function, which explains the nonlinear dependence of particle magnetization on the local magnetic field:

$$M(r,t) = M_s \cdot \mathcal{L}\left(\frac{\mu_\circ m |H(r,t)|}{k_B T}\right) \cdot \frac{H(r,t)}{|H(r,t)|} \qquad (5)$$

where $M_s$ is the saturation magnetization, $\mu_\circ$ is the magnetic permeability of free space, $m$ is the magnetic moment of an individual particle, $k_B$ is the Boltzmann constant, $T$ is the temperature in Kelvin and $\mathcal{L}$ is the Langevin function. The basic simulation produces a spatiotemporal map of the magnetization of MNPs over the FOV. These magnetization dynamics simulate the voltage signals induced in the receive coils, which are computed in the next simulation stages. To ensure consistent excitation conditions, we employ a fixed drive field amplitude throughout the entire simulation. The phantom structure from section 3.1.1 is used to construct a spatial distribution of particles. The particle concentration follows a Gaussian distribution centered in the phantom and is masked by the vessel structure to ensure particles are contained in relevant positions (see Fig.3). The FFP's position at each time step ($t$) is determined using a defined function based on time, the angular frequency of the drive field $\omega$, the radius of the circular trajectory in the xy-plane $A$, and selected trajectory patterns. To account for the z-axis, a constant of 0.05 is employed in the FFP trajectory, allowing for a maximum displacement of ±5 cm at twice the drive frequency. This modifies the FFP path to enable 3D spatial encoding in the simulated MPI system, the FFP motion is given as:

$$FFP(t) = [A\cos(\omega t), A\sin(\omega t), 0.05\sin(2\omega t)] \qquad (6)$$

The precomputed $M(r,t)$ carrying the relaxation data is loaded for all spatial locations and its time derivative dM/dt is obtained using finite differences for each component ($M_x$, $M_y$, $M_z$) across the time axis. At each voxel with non-zero particle concentration and for every time step the induced voltage signal is computed by first evaluating the distance vector between the voxel location $r$ and the instantaneous FFP position $r_{FFP}(t)$. The magnitude $|r - r_{FFP}(t)|$ is then used to calculate the field propagation factor $p(r)$ as:

$$p(r) = \frac{r - r_{FFP}(t)}{|r - r_{FFP}(t)|^3} \qquad (7)$$

which models the magnetic field's spatial decay and directional influence based on the reciprocity principle. By shifting from Cartesian axes to the Lissajous trajectory of the FFP, the induced voltage at a time $t$ $V_{lis}(t)$ given by:

$$V_{lis}(t) = \frac{dM(t)}{dt} \cdot p(r - FFP(t)) \cdot c \cdot u_{lis}(t) \qquad (8)$$

where $c$ is the local particle concentration, $u_{lis}(t)$ is the unit vector in FFP direction and $p(r - FFP(t))$ is the Lissajous propagation factor, then by integrating physical characteristics including the coil radius R, number of receive coil turns N, coil sensitivity factor α, and magnetic permeability of free space $\mu_\circ$, the system function at each spatiotemporal point is derived as:

$$SF(t) = -\frac{\mu_\circ}{4\pi} NR^2 \alpha . V_{lis}(t) \qquad (9)$$

The simulation domain consists of a 3D grid with dimensions 37 × 37 × 37, totaling 50,653 voxels. at each voxel, the system function data previously computed is stored separately for the x, y, and z receive coil directions. The model calculates the system matrix by iterating over all voxels in the 3D grid and loading time-domain system function signals from the $x$, $y$, and $z$ channels. These signals are concatenated and transformed using the Fast Fourier Transform (FFT) to extract their frequency components representing nanoparticles' nonlinear magnetic response. The resulting frequency-domain vector is flattened and saved as a column in the system matrix, indicating the voxel's whole directional response.

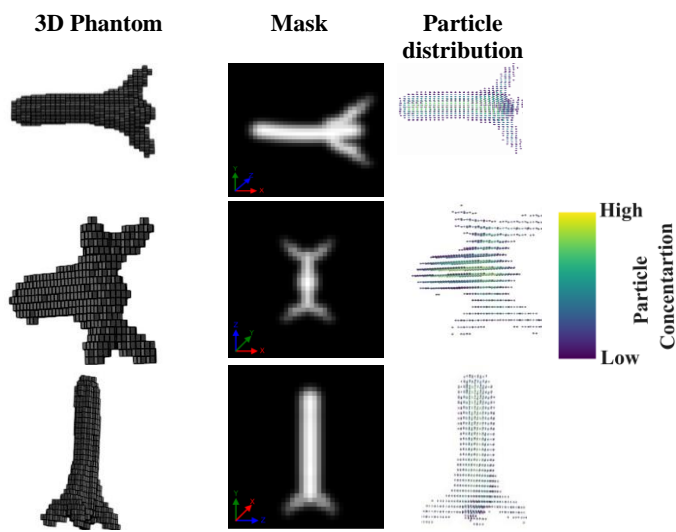

**Fig.3.** The 3D phantom structures (first column), their binary masks (second column), and simulated particle distributions (third column) are shown along three orthogonal planes: xy, xz, and yz (top to bottom). The binary masks are generated by projecting the 3D phantom onto each plane, then using a Gaussian filter and thresholding to highlight vessel-like structures. The binary masks serve as spatial priors, guiding realistic and anatomically accurate nanoparticle placement.

**2.2 CAHA implementation on the induced voltage signal**

The signals $V_{lis}(t)$ have a high-frequency content, including higher-order harmonics that are directly affected by particle distribution and system setup. To take advantage of this, we employ CAHA. This method systematically examines induced voltage signals in the x, y, and z axes and extracts the most significant harmonics to increase signal quality and reconstruction accuracy.

Each signal is transformed to the frequency domain to extract the frequency components. A Hamming window is used to reduce spectrum leakage and improve peak detection. This produces a one-sided magnitude spectrum of $V_{lis}(t)$ given as $V_d(f)$ for each $d \in \{x, y, z\}$ this magnitude spectrum is normalized to enable uniform peak detection across signals with different magnitudes.

The harmonic components are extracted by identifying local maxima in the frequency domain with a peak-finding algorithm that limits peak prominence, height, and minimum distance. We focus on peaks that correspond with odd harmonics of the fundamental frequency, $f_0$ i.e. $(3f_0, 5f_0, 7f_0, ...)$, which are known to carry the majority of spatial information due to the nonlinear Langevin behavior of nanoparticle magnetization. While even harmonics may also appear, often due to system asymmetries or signal mixing, their contribution is generally smaller and less spatially informative than the odd harmonics.

An identified harmonic frequency $f_h$ is accepted if it lies within a narrow window around an integer multiple of $f_0$ and has a sufficiently high signal-to-noise ratio (SNR), which is obtained as follows:

$$SNR_{f_h} = 20 \, log_{10} \left( \frac{|V_d(f)|}{\sigma_{noise}} \right) \tag{10}$$

where $\sigma_{noise}$ is estimated via a clustering-based approach. Instead of assuming a fixed noise region, we apply a simple Hierarchical Density-Based Spatial Clustering (HDBSCAN) to the spectrum magnitude $V_d(f)$, reshaped into a 1D feature space. We use a minimum cluster size of 5 to detect dense harmonic clusters while staying sensitive to noise. The noise cluster is chosen as the one with the lowest average magnitude, and we apply a ±10 kHz buffer to account for transition zones. A weighting factor can be applied to emphasize specific spectral regions if needed. We choose this approach because it adaptively identifies the noise without relying on fixed assumptions about the spectrum. After obtaining SNRs for each harmonic, we dynamically adjust the harmonic selection threshold based on the statistical characteristics of the SNR distribution. Particularly, the adaptive SNR threshold $T_{SNR}$ defined as the percentile $p$ of the SNR values and it's adaptively tuned based on the mean, standard deviation, and the maximum value $SNR_{max}$ of the SNR distribution. This results in a thresholding strategy that adapts to the signal quality per axis and the harmonics with $SNR_{f_h} > T_{SNR}$ are retained for reconstruction. Each analyzed axis results in multiple SNR values the average is taken for each axis.

Each retained harmonic $f_h$ is then assigned an adaptive weight $w_h$ which balances signal fidelity and the noise, the weighting incorporates a logarithmic scaling of the relative SNR:

$$w_h = \left( \gamma . log(1 + SNR_{f_h}/SNR_{max}) + (1-\gamma) . \frac{k}{N} \right) . \left( 1 + \beta . \frac{SNR_{f_h}}{SNR_{max}} \right) \tag{11}$$

Where $\gamma$ controls the trade-off between the resolution and harmonic index $k$, $\beta$ emphasizes harmonics with high SNR, and $N$ total number of harmonics. The parameter $\gamma$ is designed to reflect how spread out the SNR values are. When the SNR values vary a lot, meaning the standard deviation is high compared to the mean, then $\gamma$ increases. This helps the model give more emphasis to harmonics that stand out from the rest. On the other hand, $\beta$ becomes larger when the average SNR is close to the highest SNR value. This tells the system that most harmonics are strong and reliable, so it increases their overall influence in the weighting process. The simplified pseudocode in Algorithm 1 shows how the $\gamma$ and $\beta$ were computed.

For each direction, a harmonic set $\dot{H}_d = \{(f_h, |V_d(f)|, SNR_{f_h})\}$ is stored for further manipulation.

One of CAHA's key strengths is its directional adaptivity. Because the magnetic field shape and particle orientation can change across axes, the significant harmonic content can differ in the x, y, and z directions. CAHA addresses this by examining each direction separately, allowing harmonics to be kept even if they are powerful in one axis but weak or missing in another, as seen in Table 2. This cross-axis approach preserves all relevant spatial information during signal reconstruction.

After selecting the harmonics, we reconstruct the frequency domain signal by applying a noise floor $\varepsilon$ of 1% of the mean magnitude to unselected frequencies[24] while preserving the original phase $\varphi_k$ to stabilize the reconstruction. This eliminates sharp discontinuities. Specifically, for each direction:

$$\tilde{V}_d(f) = \begin{cases} w_h |V_d(f)| \cdot e^{-i\varphi_k}, & if\ f = f_h \in \dot{H}_d \\ \varepsilon, & other\ wise \end{cases} \quad (12)$$

---

**Algorithm 1: Adaptive Harmonic Weighting**

**Input:**
  - SNR_fh: Signal-to-noise ratio of harmonic f_h
  - SNR_max: Maximum SNR among all harmonics
  - k: Harmonic index (1, 2, ..., N)
  - N: Total number of harmonics

Step 1: Compute statistical parameters
  μ_SNR = Mean (SNR_fh for all harmonics)
  σ_SNR = Standard Deviation (SNR_fh for all harmonics)

Step 2: Compute adaptive coefficients
  γ = σ_SNR / (σ_SNR + μ_SNR) ⟶ balances log term vs. harmonic index.
  β = μ_SNR / (μ_SNR + σ_SNR) ⟶ controls SNR scaling factor.

  Bounds:
   0 ≤ γ ≤ 1
   0 ≤ β ≤ 1

Step 3: Compute harmonic weight
  w_h = (γ * log (1 + (SNR_fh / SNR_max)) + (1 - γ) * (k / N))
     * (1 + β * (SNR_fh / SNR_max))

**Output:**
  - w_h : Adaptive weight for harmonic f_h

---

**Table 2.** The adaptive selection and parameter tuning results using CAHA at a drive frequency of 75 kHz

| Axis | $Avg. f_h$(kHz) | $Avg\ SNR_{f_h}$ (dB) | $T_{SNR}$(dB) | $\gamma$ | $\beta$ | $p$ |
|---|---|---|---|---|---|---|
| x | 76.67 | 4.59 | 3.7 | 0.89 | 0.55 | 77.11 |
| x | 153.33 | 4.35 | 3.7 | 0.89 | 0.55 | 77.11 |
| x | 306.67 | 7.26 | 3.7 | 0.89 | 0.55 | 77.11 |
| y | 76.67 | 0.74 | 0.20 | 0.92 | 0.55 | 75.77 |
| z | 76.67 | 5.20 | 1.36 | 0.92 | 0.54 | 76.01 |
| z | 230 | 1.70 | 1.36 | 0.92 | 0.54 | 76.01 |

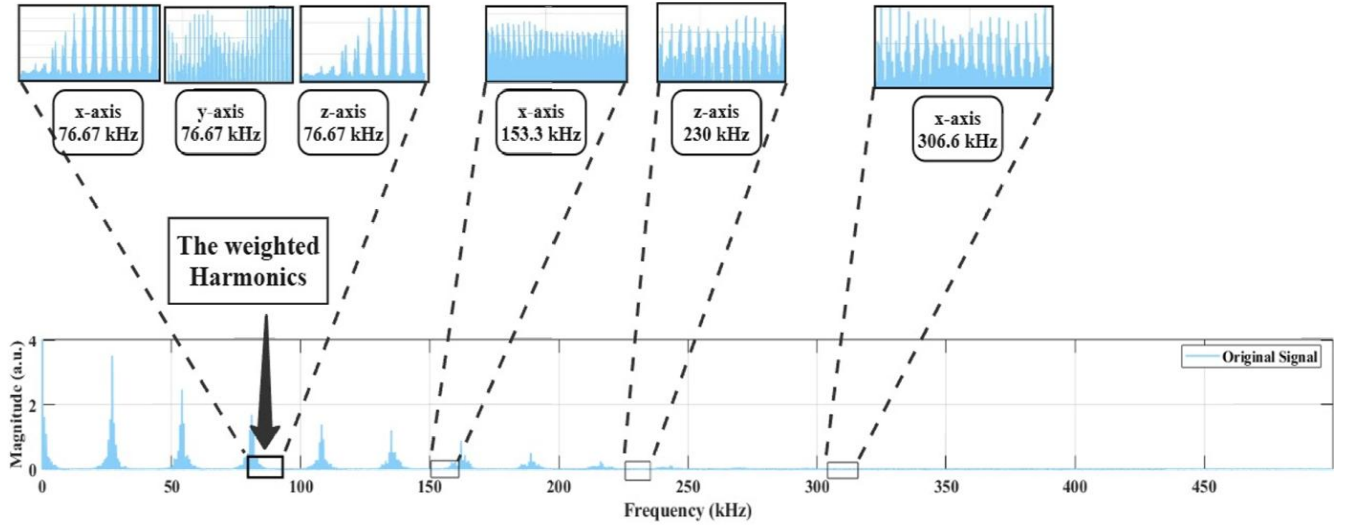

**Fig.4.** The frequency-domain spectrum of the induced voltage signal simulated at a drive frequency of 75 kHz. Highlighted black squares respectively to indicate the frequency regions selected by the CAHA method based on harmonic strength and signal quality. These selected harmonics were adaptively weighted across the x, y, and z axes, as shown in the zoomed-in waveforms, to generate a reconstructed signal used for image reconstruction.

### 2.3. Demonstration of CAHA on Open MPI Dataset

The CAHA method is also tested on the Open MPI dataset and evaluated using three types of phantoms: the resolution phantom, the shape phantom, and the concentration phantom. Unlike the simulated data, where the drive frequency was varied and the analysis performed in each axis individually, here, the drive frequency is fixed at 2.5 MHz for all signals. The system matrix data used for reconstruction with all phantoms is calibration No.3 with 19×19×19 grid size, 38× 38 × 19 FOV and measured with Perimag tracer, the detailed calibration metadata is reported in [25]. The adaptive parameter settings used for harmonic selection with CAHA are provided in Table 3. CAHA adaptively focuses on the mid-frequency band, primarily within the 200–300 kHz, as seen in Fig.5, ensuring a more stable and accurate reconstruction.

**Table 3.** The adaptive selection parameters for the Open MPI dataset using CAHA.

| Phantom | Axis | $Avg.f_h$(kHz) | $Avg.SNR_{f_h}$ (dB) | $T_{SNR}$(dB) | $\gamma$ | $\beta$ | $p$ |
|---|---|---|---|---|---|---|---|
| Resolution | x | 289.10 | 14.27 | 8.87 | 0.83 | 0.57 | 82.08 |
|  | y | 290.12 | 16.06 | 8.87 | 0.83 | 0.57 | 82.08 |
|  | z | 267.23 | 15.02 | 8.87 | 0.83 | 0.57 | 82.08 |
| Shape | x | 292.20 | 15.11 | 9.79 | 0.82 | 0.58 | 75.77 |
|  | y | 281.68 | 16.09 | 9.79 | 0.82 | 0.58 | 76.01 |
|  | z | 268.85 | 15.70 | 9.79 | 0.82 | 0.58 | 76.01 |
| Concentration | x | 290.71 | 14.72 | 8.93 | 0.83 | 0.57 | 82.10 |
|  | y | 288.59 | 15.28 | 8.93 | 0.83 | 0.57 | 82.10 |
|  | z | 261.92 | 15.12 | 8.93 | 0.83 | 0.57 | 82.10 |

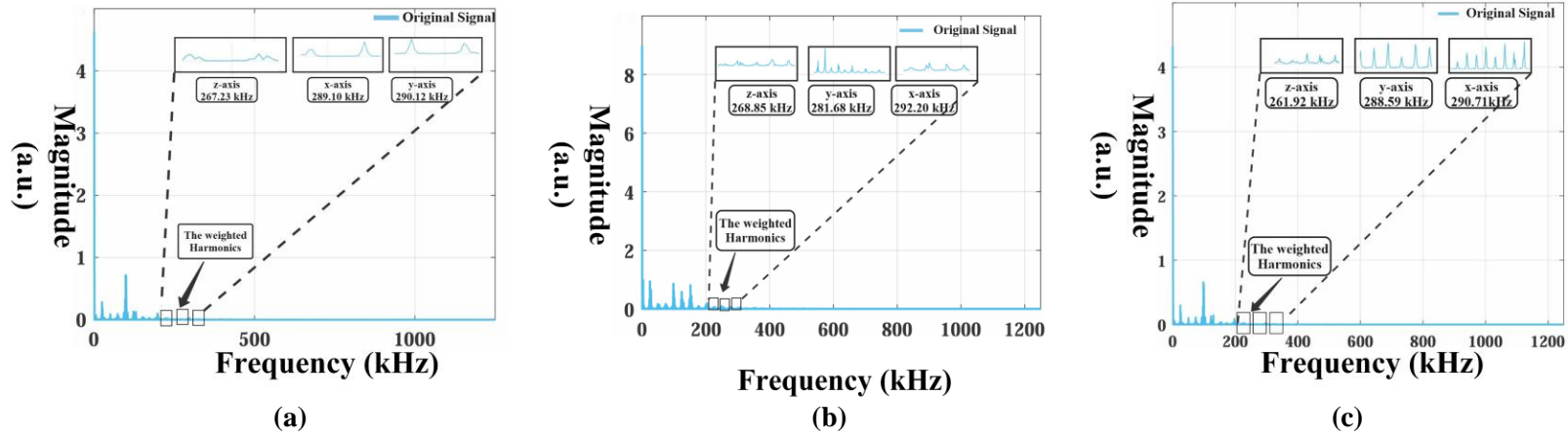

**Fig. 5.** Frequency-domain representation of the induced voltage signals from the Open MPI dataset for three different phantoms: (a) resolution phantom, (b) shape phantom, and (c) concentration phantom. The weighted harmonics, used to reconstruct the signal for image reconstruction, are ordered sequentially from left to right based on their frequency. The zoomed-in insets illustrate dominant harmonics across the x, y, and z axes, capturing the unique characteristics of each phantom.

**2.4. Image reconstruction**

To assess the robustness of the reconstructed signal obtained through CAHA, we evaluated its effectiveness using different image reconstruction techniques. Our primary comparison is based on the Kaczmarz method [5], where we contrast images reconstructed from the original induced voltage signal with those reconstructed from the CAHA-processed signal. To further validate the generalizability of the CAHA-enhanced signal, we also implemented additional reconstruction methods beyond Kaczmarz, such as Fast iterative shrinking-thresholding algorithm (FISTA)[26][27], Tikhonov[5], least square matrix decomposition (LSQR)[28][29][30], and total variation minimization (TV)[31][30]. This allows us to examine whether the improvements introduced by CAHA are consistent across different reconstruction strategies and not limited to a specific solver.

Each reconstruction method relies on important parameters that influence the image quality. For the Kaczmarz method, the regularization factor $\lambda = 0.75$ helps stabilize the iterative updates and control overfitting to noise. In FISTA, the regularization weight $w = 0.1$ is used, and the maximum number of iterations $= 1000$, with internal handling of convergence using a computed Lipschitz constant. Tikhonov regularization also uses a single regularization parameter $\lambda_T = 0.45$ to balance fidelity and smoothness in its closed-form solution.

The LSQR method includes a damping parameter $d = 0.45$, acting similarly to Tikhonov, and an iteration limit $= 100$ to control convergence. For the TV method, a regularization weight $w_{TV}$ of $0.1$ and iterations $= 100$ are used to enforce edge-preserving smoothing. A small epsilon $= 1e^{-8}$ is also used internally to ensure stable gradient calculations.

**Table 4.** Parameters used in the simulation model.

| Parameter | Physical Unit | Value | Description |
|---|---|---|---|
| Drive Frequency ($D_f$) | kHz | 20 to 85 | The fundamental frequency used to generate the drive field in MPI. |
| Gradient Field Strength ($H_s$) | T/m | 5 | Defines the spatial resolution and sensitivity of MPI by determining the field-free region. |
| Focusing Field Strength ($H_F$) | mT | 3 | Adjusts the field-free point to enhance image quality. |
| Magnetic Moment (m) | A·m² | $9.6 \times 10^{-23}$ | Represents the response of the nanoparticles to the external field. |
| Induced Voltage Signal ($V_{x,y,z}$) | mV | Computed dynamically | The voltage signal induced in the receiver coil due to nanoparticle magnetization changes. |
| Field of view (FOV) | mm | 50× 50 × 50 | Defines the 3D imaging volume where the MNPs are spatially encoded. |
| Noise Level | Arbitrary Units | 0.1(normalized) | Represents background noise in the detected signal. |
| SNR Threshold ($T_{SNR}$) | dB | Adaptive | Used in filtering harmonics based on the SNR. |
| Coil Sensitivity | V/A·m² | Defined per coil | Determines how efficiently the receiver coil detects induced signals. |
| Spatial Grid Resolution | mm | 2.56 | Defines the discretization of the phantom and field simulation. |
| Drive Field Amplitude | mT | 10 (maximum) | The magnitude of the oscillating magnetic field is applied to excite the nanoparticles. |
| Temperature | K | 293 | Affects the thermal energy and magnetic response of nanoparticles. |
| Boltzmann Constant | J/K | $1.38 \times 10^{-23}$ | Governing nanoparticle magnetization statistics. |
| Nanoparticle Radius | nm | 25 | Determines the size of the nanoparticles, influencing their magnetic properties. |
| Brownian / Néel's Relaxations | s | $10^{-3}$ / $10^{-7}$ | Describes how nanoparticles' magnetization rotates due to Brownian motion / Describes how nanoparticles' magnetization flips internally due to thermal energy. |

**Table 5.** Some terms used in this work

| Term | Definition |
|------|------------|
| xyO | The xy view for the image reconstructed with the original signal. |
| xyR | The xy view for the image reconstructed with the reconstructed signal. |
| xzO | The xz view for the image reconstructed with the original signal. |
| xzR | The xz view for the image reconstructed with the reconstructed signal. |
| yzO | The yz view for the image reconstructed with the original signal. |
| yzR | The yz view for the image reconstructed with the reconstructed signal. |

## 3. Results

### 3.1. For the simulated model

The reconstructed images from the simulated numerical phantom showed notable improvements when using the reconstructed induced voltage signals compared to the original signals. The images from the original signals showed lower resolution and more noise. This enhancement was especially evident in regions with weaker signals, where the noise reduction was most beneficial.

The results were validated by the Full Width at Half Maximum (FWHM) plots, which showed narrower FWHM values for the images processed with the reconstructed signals, confirming the improvement in image sharpness. Across all axes x, y, and z, the adaptive signal processing helped achieve more consistent and high-quality reconstructions, highlighting the value of the cross-axis harmonic analysis in boosting resolution and minimizing noise.

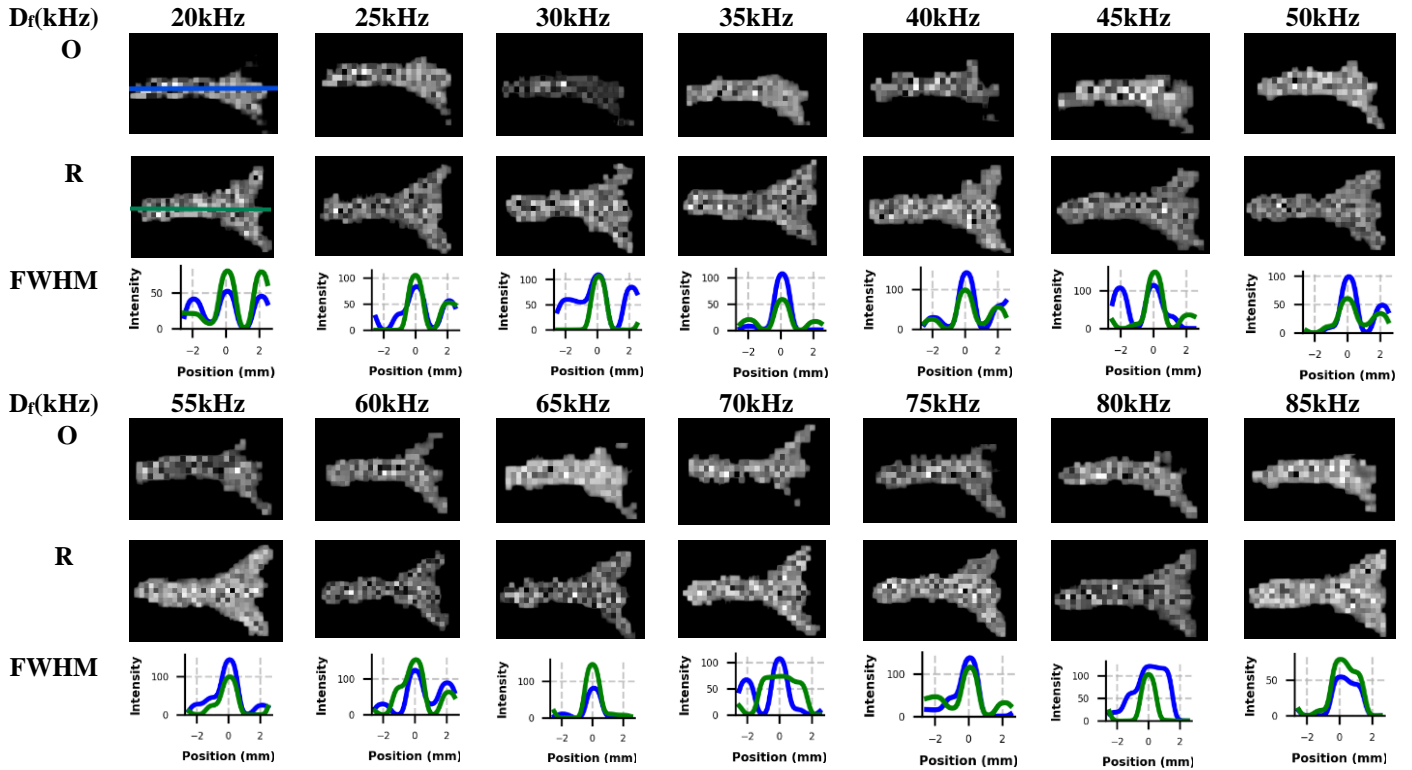

**Fig.6**. Reconstructed images using Kaczmarz method with corresponding FWHM profiles at various drive frequencies ($D_f$) for both original (O) and reconstructed (R) induced voltage signals. Each image shows the reconstruction of the simulated numerical phantom composed of several voxels in the xy view, with the drive frequency values ranging from 20 kHz to 85 kHz. Below each image pair, the associated FWHM plot is displayed, with the original induced voltage signal in blue and the reconstructed induced voltage signal in green. The FWHM graphs depict the intensity profile along the position (mm) for both signals at every drive frequency.

As depicted in Fig. 6, at lower drive frequencies (20-35 kHz), reconstructed images (R) exhibit comparably clear and consistent representations of the phantom, but the spatial resolution of the images reconstructed with the original induced voltage signal (O) is less detailed. The difference between (O) and (R) images becomes increasingly noticeable as the drive frequency increases (40-85 kHz). The original induced voltage signal (O) exhibits blurring and loss of fine details for higher frequencies, notably around the borders of the phantom's features. This is owing to the higher frequencies introducing more high-frequency components, which are not captured accurately in the original signal. On the other hand, the reconstructed images (R), which are processed by CAHA, reveal higher spatial resolution with sharper and more defined structures. The higher-frequency components in the signal are better maintained and depicted in these images, making them substantially more detailed and precise.

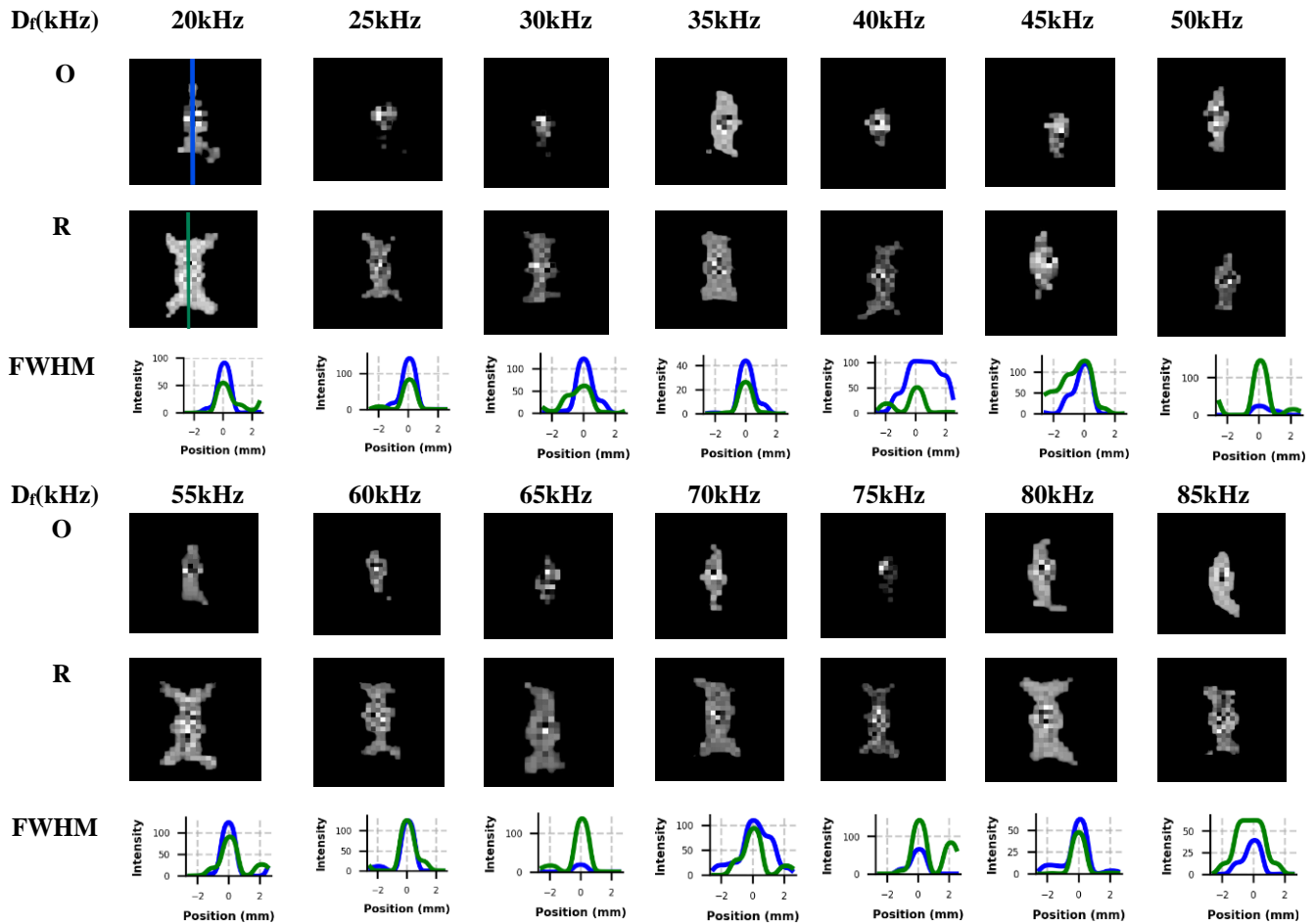

**Fig.7.** Reconstructed images of the xz view of the simulated numerical phantom with corresponding FWHM profiles at various drive frequencies ($D_f$) for both original (O) and reconstructed (R) induced voltage signals. The layout and interpretation are identical to Fig. 6.

As shown in Fig.7 and 8, for most of the drive frequencies, the FWHM profiles of the reconstructed images R appear to be narrower than those of the reconstructed images (O), showing that cross-axis harmonic analysis improves the spatial resolution. However, an interesting exception occurs at the maximum frequency, 85 kHz, where the FWHM for images (O) becomes narrower than that of the reconstructed images (R), although the difference is a little as in Fig. 6 and 8 with slightly large difference as in Fig. 7, at this frequency, the original signal retains more precise intensity profiles, whereas the CAHA used for the reconstructed images (R) seems to produce some degree of blurring or artifacts, leading to a broader FWHM. This trend can be explained by higher frequencies capturing more fine-grained signal characteristics, and relaxation effects at this frequency are constraining the nanoparticles' response.

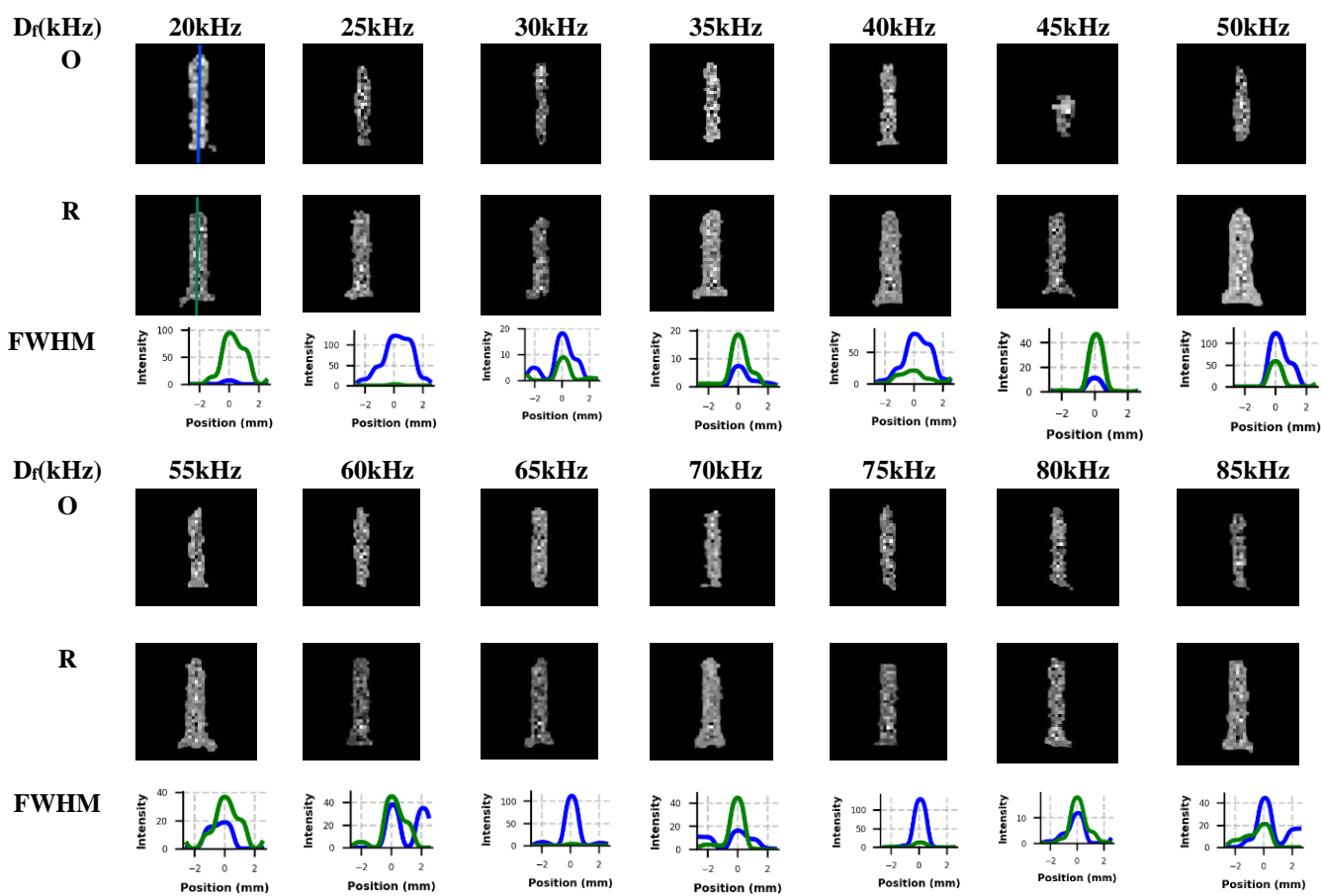

**Fig.8.** Reconstructed images of the yz view of the simulated numerical phantom with corresponding FWHM profiles at various drive frequencies ($D_f$) for both original (O) and reconstructed (R) induced voltage signals. The layout and interpretation are identical to Fig. 6 and Fig.7.

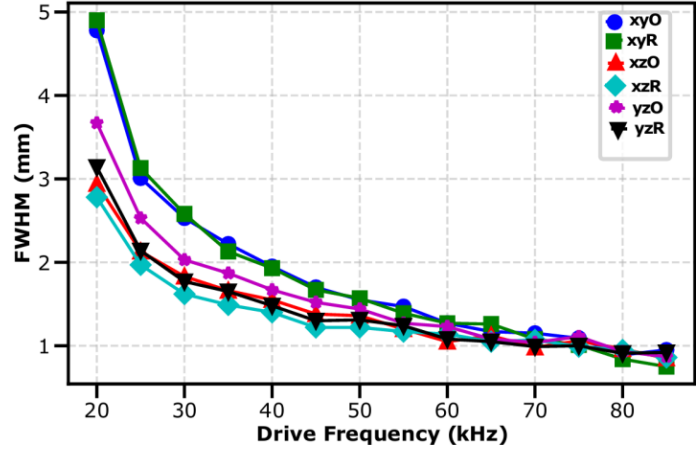

**Fig. 9.** FWHM versus drive frequency for the images reconstructed with the original induced voltage signal and with the reconstructed induced voltage signal for different views.

Regarding the resolution of features, the reconstructed images (R) in the xy, xz, and yz views exhibit markedly improved spatial resolution and greater detail than the original images (O), with sharper and more precise definitions of these features. This is particularly apparent in the areas enclosed by the red boxes, as shown in Fig. 10, where the delineation of the phantom's characteristics is significantly more clearly defined in the reconstructed images (R).

**View    O          R**

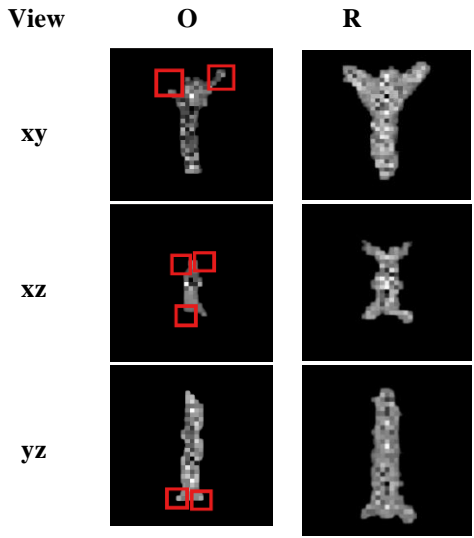

xy

xz

yz

**Fig.10.** A comparison of feature resolution for images reconstructed at a drive frequency of 55 kHz. Each view (xy, xz, yz) shows a pair of images: the left image in each pair is reconstructed from the original signal (O), and the right image is reconstructed from the CAHA-processed signal (R). Red boxes highlight specific regions where the improvement in spatial resolution and feature delineation is most apparent in the (R) images compared to the (O) images.

**Table 6.** FWHM values against the drive frequency for different reconstructed image views.

|            | FWHM (mm) |      |      |      |      |      |
|------------|-----------|------|------|------|------|------|
| $D_f$ (kHz) | xyO | xyR | xzO | xzR | yzO | yzR |
| 20         | 4.78 | 4.90 | 2.94 | 2.78 | 3.67 | 3.14 |
| 25 | 3.01 | 3.13 | 2.13 | 1.97 | 2.53 | 2.14 |
| 30 | 2.53 | 2.58 | 1.83 | 1.62 | 2.03 | 1.77 |
| 35 | 2.22 | 2.13 | 1.66 | 1.49 | 1.87 | 1.65 |
| 40 | 1.95 | 1.93 | 1.55 | 1.40 | 1.67 | 1.48 |
| 45 | 1.70 | 1.67 | 1.38 | 1.22 | 1.52 | 1.30 |
| 50 | 1.55 | 1.57 | 1.36 | 1.22 | 1.44 | 1.31 |
| 55 | 1.47 | 1.39 | 1.20 | 1.17 | 1.27 | 1.24 |
| 60 | 1.27 | 1.27 | 1.05 | 1.14 | 1.23 | 1.08 |
| 65 | 1.17 | 1.26 | 1.12 | 1.05 | 1.08 | 1.05 |
| 70 | 1.15 | 1.08 | 0.98 | 1.07 | 1.04 | 0.99 |
| 75 | 1.09 | 1.01 | 1.06 | 0.99 | 1.11 | 1.00 |
| 80 | 0.89 | 0.84 | 0.96 | 0.95 | 0.93 | 0.91 |
| 85 | 0.92 | **0.80** | 0.85 | 0.90 | 0.87 | 0.92 |

The findings in Table 6 highlight how frequency-dependent signal encoding directly impacts the point spread function with a clear trend emerging that FWHM values get lower across the reconstructed image views as the driving frequency increases. For instance, in the xyR view, the FWHM decreases from 4.90 mm at 20 kHz to at least 0.80 mm at 85 kHz. A similar pattern is evident in other views, with all reconstructions achieving sub-millimeter resolution at the highest frequencies. Comparatively, across most drive frequencies, the R views tend to have marginally lower FWHM values than their corresponding O views. The differences, although not drastic, become more evident at lower frequencies, where the reconstructions more effectively compensate for the inherently poorer resolution at lower drive frequencies.

To further assess the quality of both reconstructed image groups, we calculated the image metrics: Peak Signal-to-Noise Ratio (pSNR), normalized Root Mean Square Error (nRMSE), and Structural Similarity Index (SSIM). All these metrics are computed as follows:

$$pSNR = 10.\log_{10}\left(\frac{\left(max(I_{ref})\right)^2}{\frac{1}{N}\sum_{i=1}^{N}\left(I_{rec}(i)-I_{ref}(i)\right)^2}\right) \quad (13)$$

where $I_{ref}$, $I_{rec}$ and N denote the reference, reconstructed images, and the total number of pixels in the image, respectively. The nRMSE was calculated as:

$$nRMSE = \frac{\sqrt{\frac{1}{N}\sum_{i=1}^{N}\left(I_{rec}(i)-I_{ref}(i)\right)^2}}{max(I_{ref})-min(I_{ref})} \quad (14)$$

which normalizes the reconstruction error relative to the dynamic range of the reference image. SSIM was used to capture structural fidelity and is defined as:

$$SSIM = \frac{(2\mu_x\mu_y + C_1)(2\delta_{xy} + C_2)}{(\mu_x^2+\mu_y^2+ C_1)(\delta_x^2+\delta_y^2+ C_2)} \quad (15)$$

Where $\mu, \delta^2$ and $\delta_{xy}$ represent local means, variances, and covariance, respectively, while $C_1$ and $C_2$ are small constants to stabilize the division when denominators are close to zero.

For fair comparison, the reference image was a high-resolution ground truth generated by upscaling the phantom with an 8× for the three views xy, xz and yz. All metrics were computed across the full image domain corresponding to the reconstructed field-of-view. Both reference and reconstructed images were resized to identical dimensions and normalized to the [0, 1] range before metric computation. Implementation was performed in Python using the scikit-image library for pSNR and SSIM, and a custom function for nRMSE.

Table 7 shows the values of these metrics for both groups. Generally, the reconstructed signal R improves image quality over the original O, particularly at lower frequencies where signal degradation is more pronounced. The most noticeable enhancements are observed in the xy view; this consistently high performance across metrics indicates that reconstruction is pretty effective in this view. One plausible explanation is that the receiver coil sensitivity is improved in typical MPI systems, and the drive and selection fields are stronger in both the x and y directions. The reconstruction can utilize richer spatial information since the encoding in these lateral directions is usually stronger.

Because the selection field gradient and receiver coil sensitivity are typically weaker along the z-direction, encoding along this axis is generally less effective. As a result, fine details are harder to capture, which often leads to increased image blurring and ambiguity. While the xz view benefits from signal reconstruction, its improvements are not as significant as the xy view. For instance, at 85 kHz, the nRMSE falls to 0.05 and the PSNR reaches a peak of 29.70 dB, indicating some enhancement. Around 55 kHz, the SSIM also peaks at 0.91, indicating improved structural preservation. However, these improvements are less consistent and reliable than those seen in the xy view, likely due to the z-axis's lower resolution and limited encoding strength.

**Table 7.** The evaluation metrics for different reconstructed views using the original induced voltage and reconstructed signals.

| | xyR | | | xyO | | | xzR | | | xzO | | | yzR | | | yzO | | |
|---|---|---|---|---|---|---|---|---|---|---|---|---|---|---|---|---|---|---|
| $D_f$ (kHz) | p* (dB) | n* | s* | p (dB) | n | s | p (dB) | n | s | p (dB) | n | s | p (dB) | n | s | p (dB) | n | s |
| 20 | 15.49 | 0.09 | 0.78 | 13.26 | 0.19 | 0.50 | 24.5 | 0.20 | 0.88 | 21.80 | 0.29 | 0.59 | 16.05 | 0.32 | 0.81 | 11.80 | 0.20 | 0.59 |
| 25 | 17.55 | 0.08 | 0.77 | 16.7 | 0.16 | 0.38 | 22.47 | 0.28 | 0.73 | 14.41 | 0.38 | 0.45 | 18.01 | 0.30 | 0.65 | 11.80 | 0.30 | 0.55 |
| 30 | 19.19 | 0.04 | 0.77 | 17.73 | 0.25 | 0.50 | 22.46 | 0.29 | 0.67 | 14.82 | 0.39 | 0.40 | 17.96 | 0.28 | 0.53 | 11.80 | 0.30 | 0.40 |
| 35 | 20.43 | 0.06 | 0.75 | 19.77 | 0.15 | 0.58 | 22.35 | 0.26 | 0.85 | 14.54 | 0.30 | 0.51 | 18.39 | 0.26 | 0.56 | 11.80 | 0.30 | 0.40 |
| 40 | 22.70 | 0.03 | 0.79 | 20.01 | 0.21 | 0.56 | 22.50 | 0.33 | 0.66 | 18.64 | 0.38 | 0.43 | 18.95 | 0.24 | 0.51 | 14.45 | 0.29 | 0.46 |
| 45 | 23.79 | 0.04 | 0.81 | 20.78 | 0.20 | 0.59 | 23.04 | 0.37 | 0.58 | 19.02 | 0.26 | 0.46 | 19.09 | 0.22 | 0.67 | 15.21 | 0.26 | 0.41 |
| 50 | 23.14 | 0.04 | 0.89 | 21.96 | 0.14 | 0.59 | 23.01 | 0.39 | 0.56 | 18.37 | 0.24 | 0.49 | 18.76 | 0.20 | 0.61 | 20 | 0.26 | 0.45 |
| 55 | **24.76** | 0.02 | **0.94** | 22.4 | 0.14 | 0.55 | 23.25 | 0.04 | **0.91** | 20.31 | 0.21 | 0.54 | 19.07 | 0.08 | **0.89** | 20.62 | 0.21 | 0.67 |
| 60 | 24.58 | 0.02 | 0.86 | 21.74 | 0.09 | 0.56 | 23.06 | 0.16 | 0.73 | 20.84 | 0.29 | 0.47 | 18.67 | 0.16 | 0.64 | 20.27 | 0.19 | 0.53 |
| 65 | 24.93 | 0.03 | 0.88 | 22 | 0.12 | 0.58 | 23.10 | 0.14 | 0.81 | 20.99 | 0.41 | 0.40 | 18.94 | 0.14 | 0.65 | 21.23 | 0.20 | 0.49 |
| 70 | 25.30 | 0.03 | 0.87 | 22.78 | 0.11 | 0.57 | 23.06 | 0.12 | 0.85 | 20.87 | 0.36 | 0.50 | 19.05 | 0.12 | 0.67 | 20.67 | 0.18 | 0.51 |
| 75 | 25.25 | 0.03 | 0.85 | 22.95 | 0.10 | 0.57 | 23.20 | 0.09 | 0.79 | 15.70 | 0.41 | 0.39 | 18.81 | 0.17 | 0.55 | 20.59 | 0.11 | 0.54 |
| 80 | 24.40 | 0.03 | 0.83 | 22.87 | 0.14 | 0.60 | 23.30 | 0.07 | 0.84 | 20.53 | 0.19 | 0.65 | 18.75 | 0.19 | 0.50 | 21.37 | 0.10 | 0.57 |
| 85 | 24.22 | **0.01** | 0.84 | 22.09 | 0.15 | 0.59 | **29.70** | 0.05 | 0.87 | 20.47 | 0.16 | 0.60 | **29.70** | 0.10 | 0.77 | 20.72 | 0.16 | 0.56 |

*p, n, and s are the pSNR, nRMSE and SSIM, respectively.

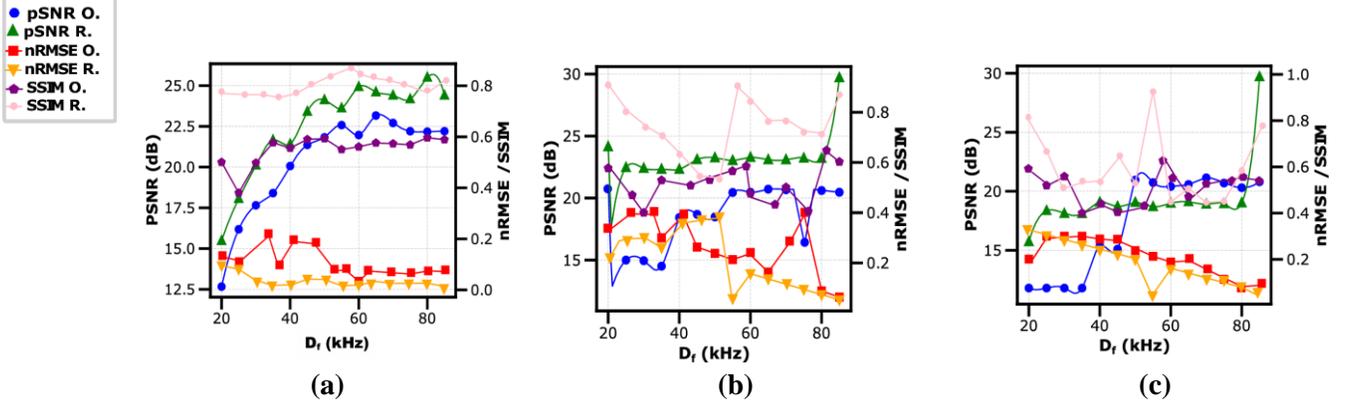

**Fig.11.** Metrics trend for the reconstructed images (R) versus original (O) across drive frequencies ($D_f$) for (a) xy, (b) xz, and (c) yz views.

From the studied range, 85 kHz emerges as the optimal drive frequency, providing the best overall balance of spatial resolution and image quality throughout the evaluated views. At this frequency, the xyR view reaches a sub-millimeter FWHM, showing good spatial resolution, and has the lowest nRMSE (0.01) of any view, implying minimal reconstruction errors. While the PSNR (24.22 dB) and SSIM (0.84) are not the absolute greatest values recorded, they are consistently powerful and well-balanced, especially when combined with the excellent FWHM. In challenging views, xzR and yzR, 85 kHz still results in measurable improvements, albeit with slightly higher nRMSE and lower SSIM, most likely due to worse encoding in the z-direction (see Supplementary Material Section 3). These results support the interpretation that higher drive frequencies enhance the induced voltage signal's harmonic content, improving spatial encoding and image quality, though the relationship is not strictly linear and depends on the view and axis-specific encoding sensitivities.

Alongside 85 kHz, the 55 kHz drive frequency also demonstrates impressive performance. It attains the highest SSIM values in the xyR, xzR, and yzR views, suggesting strong structural fidelity, and achieves a notable PSNR of 24.76 dB in the xyR view, outperforming 85 kHz in this aspect. Additionally, it shows low reconstruction errors, as in Table 7.

To clarify the choice of drive frequency, we evaluated four key metrics simultaneously: FWHM, pSNR, SSIM, and nRMSE. Rather than relying on a single metric, we applied a Pareto-style trade-off approach[32], where a frequency is considered optimal if it improves at least one metric without significantly compromising the others. Using this multi-metric perspective, two frequencies stand out and appears on the Pareto front. At 85 kHz, the overall performance is most balanced, achieving sub-millimeter resolution while maintaining strong pSNR and SSIM, and low nRMSE across all views, making it a robust choice. In contrast, 55 kHz achieves the highest structural similarity (SSIM = 0.94 in xyR) and competitive PSNR with low NRMSE, indicating it best preserves structural details even though its resolution is slightly lower than 85 kHz. Figure 12 shows the Pareto analysis plots based on these metrics, illustrating how the different frequencies compare across all evaluation criteria.

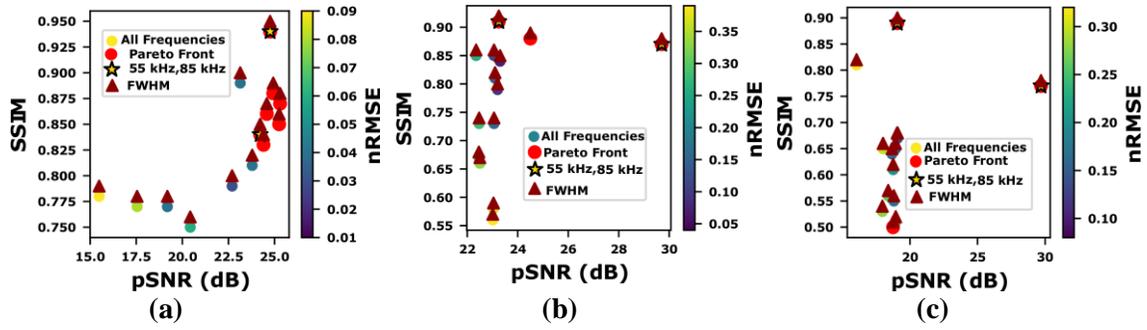

**Fig. 12.** Pareto analysis of drive frequencies based on multiple image quality metrics, each plot shows FWHM, pSNR , SSIM, and nRMSE for the image reconstructed using the reconstructed signals (a) xyR, (b) xzR, and (c) yzR view. All the metric values in Tables 6 and 7 are used in this analysis.

### 3.2. Other image reconstruction methods

The induced voltage signal reconstructed using the CAHA method and the original signals from the simulation model are then used for image reconstruction via the predefined algorithms in Section 2.4, as shown in Figure 13. Importantly, while all methods effectively recover the overall vessel geometry, there are clear visual differences in sharpness, smoothness, and background noise in the reconstructions.

Table 8 presents the metrics to quantify image quality; the TV-based reconstruction outperformed all others across all views in the case of the reconstructed signal (R), achieving the highest pSNR values, the lowest nRMSE, and the highest SSIM, which indicate superior fidelity, noise suppression, and structural preservation. LSQR and FISTA follow, yielding moderately good reconstructions, while the Tikhonov method performs poorly in the case of reconstructed signal (R), exhibiting visibly degraded structures and low quantitative scores. Interestingly, Tikhonov performs better in the case of the original signal (R), showing higher pSNR and SSIM and lower nRMSE compared to the reconstructed signal (R). Overall, the results validate that combining CAHA signal processing with robust regularization results in high-quality reconstructions from MPI signals.

| View / Method | xy | xz | yz |
| --- | --- | --- | --- |
| Tikhonov (R) | | | |
| Tikhonov (O) | | | |
| FISTA (R) | | | |

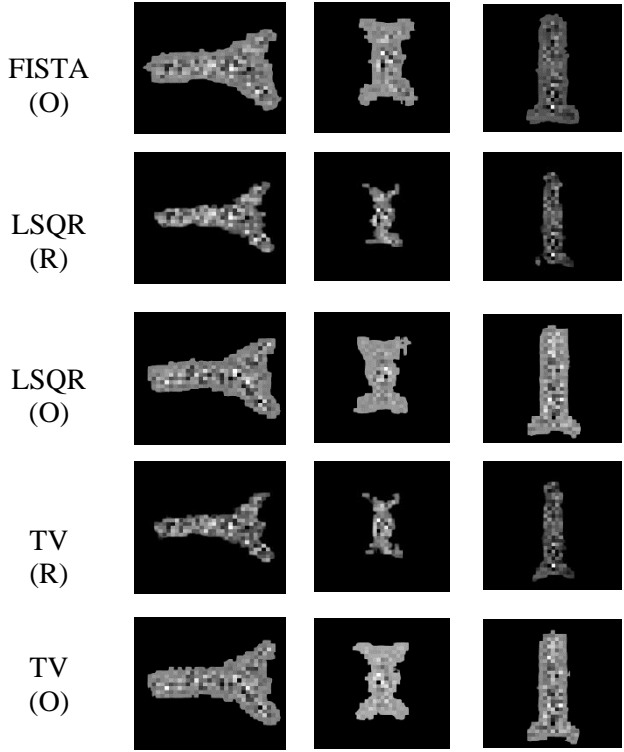

**Fig.13.** Reconstructed phantom images at a drive frequency of 55 kHz using the induced voltage signal processed with CAHA (R) and the corresponding images reconstructed with the original signal(O) with various reconstruction methods: Tikhonov, FISTA, LSQR, and TV. The reconstructions are shown across three orthogonal views (xy, xz, yz) to highlight the spatial structure and continuity of the 3D vessel phantom.

**Table 8.** The evaluation metrics for the other reconstruction methods are performed using the reconstructed induced voltage signal (R) and the original signal (O) at a drive frequency of 55 kHz.

| Metrics | pSNR | | | nRMSE | | | SSIM | | |
|---|---|---|---|---|---|---|---|---|---|
| Method/View | xy | xz | yz | xy | xz | yz | xy | xz | yz |
| Tikhonov | | | | | | | | | |
| (R) | 11.44 | 10.15 | 9.58 | 1.45 | 2.30 | 2.92 | 0.54 | 0.50 | 0.53 |
| (O) | 14.70 | 13.35 | 13.21 | 0.99 | 1.59 | 1.92 | 0.64 | 0.74 | 0.72 |
| FISTA | | | | | | | | | |
| (R) | 23.69 | 24.58 | 25.00 | 0.35 | 0.44 | 0.49 | 0.89 | 0.95 | 0.90 |
| (O) | 15.17 | 14.11 | 19.59 | 0.94 | 1.46 | 0.92 | 0.74 | 0.78 | 0.81 |
| LSQR | | | | | | | | | |
| (R) | 24.84 | 25.37 | 28.56 | 0.32 | 0.40 | 0.33 | 0.91 | 0.95 | 0.95 |
| (O) | 14.29 | 14.37 | 14.07 | 1.04 | 1.42 | 1.74 | 0.70 | 0.80 | 0.77 |
| TV | | | | | | | | | |
| (R) | **37.91** | **35.50** | **31.80** | **0.09** | **0.11** | **0.14** | **0.97** | **0.96** | **0.95** |
| (O) | 14.40 | 13.91 | 14.63 | 1.03 | 1.49 | 1.63 | 0.70 | 0.81 | 0.77 |

### 3.3. For Open MPI Data

To confirm the proposed CAHA method's efficiency, we tested it on the publicly available Open MPI dataset. This configuration allows us to explore how CAHA behaves under high drive field conditions and how it affects the induced voltage signal reconstructed from different phantoms.

As in Fig. 14, the method shows considerable increases in all image quality measures for the resolution and shape phantoms. In both cases, pSNR and SSIM increase significantly, whereas nRMSE is lowered or maintained. Most importantly, the FWHM is reduced dramatically, showing improved spatial resolution and delineation of true fine structures, which are crucial in MPI for accurately resolving small or closely spaced features.

Although the reconstructed image R has a slightly larger FWHM in the case of the concentration phantom (1.20 mm vs. 1.00 mm in O), it achieves a greater pSNR and SSIM, indicating improved overall image fidelity and structural consistency. This apparent discrepancy most likely represents a common trade-off between resolution and signal in MPI. In this scenario, CAHA most certainly reduces noise and enhances the visibility of low-concentration regions, increasing structural accuracy across the image at the expense of a little loss in peak sharpness. Such behavior is expected in high-frequency MPI settings, where keeping relevant signal content and structural integrity can be more important than maximizing point resolution.

In the concentration phantom case, we observed that FWHM values worsened slightly while pSNR and SSIM improved. This outcome reflects a resolution–noise trade-off: CAHA suppresses low-SNR harmonics, thereby reducing noise and enhancing structural similarity to the reference image, at the cost of minor resolution loss. Importantly, the increase in PSNR and SSIM demonstrates that, despite the broader point spread, the reconstructed images exhibit higher fidelity and better perceptual quality. This indicates that CAHA prioritizes noise suppression and global consistency while maintaining resolution within an acceptable range, a balance commonly observed in advanced image-processing frameworks.

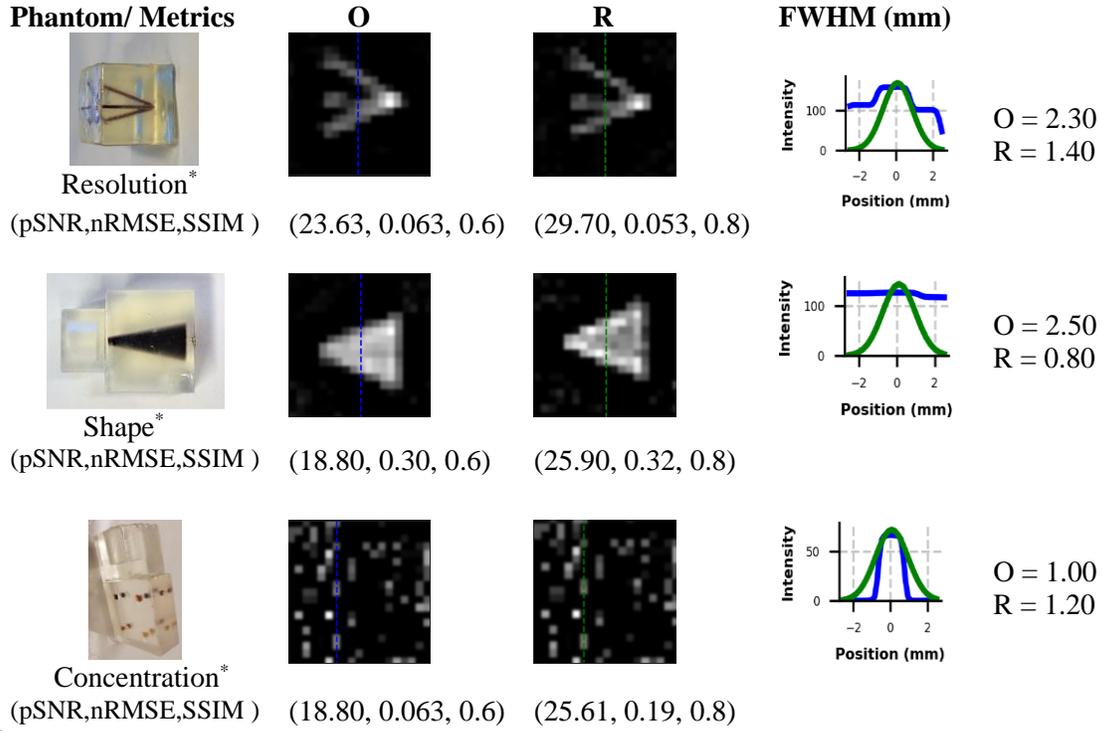

**Fig.14.** The image reconstructed for the Open MPI Dataset using the original signal O and reconstructed signal R for the xy view, images are given alongside their corresponding FWHM values, whereas the metrics pSNR, nRMSE and SSIM are provided under each image. All the images are reconstructed with Kaczmarz method.

**Table 9.** Comparison of MPI image reconstructed quality metrics across different State-of-the-art methods.

| Study | FWHM (mm) | pSNR (dB) | nRMSE | SSIM | Notes |
|---|---|---|---|---|---|
| Drive Field Optimization[16] | 1.6 →0.9. | - | - | - | Experimental optimization at 70 kHz and 5 mT/$\mu°$. |
| Simulation based system matrix[33] | - | 20.5250 | 0.0906 | 0.8668 | Tikhonov-regularized using simulation data |
| Weighted harmonics[20] | - | signal-to-artifact ratio (SAR) ↑, 2.5×. | ↓88%. | ↑48%. | simulations + experimental data. |
| 3D MPI simulation model[19] | - | 36.49 | - | 0.881 | simulation + the Open MPI data. |
| SNR-Based frequency selection[22] | 2.40, 2.48, and 2.85 | - | - | - | simulation + real data. |
| **Ours** | **0.8** at 85 kHz in xyR (Simulation) **0.8** (Open MPI data) | 29.7 at 85 kHz in xzR and yzR. 29.7 (Open MPI data). **37.91** at 55 kHz using total variation. | **0.01** at 85 kHz in xyR (simulation). **0.053** (Open MPI data). | **0.94** at 55 kHz in xyR (simulation). **0.8** (Open data). | simulation + the Open MPI dataset. Optimum drive frequency 55 & 85 kHz at 10mT drive field amplitude. |

## 3.4. Ablation study
### 3.4.1. Using different hyperparameters

To better understand the effect of parameter choices on image reconstruction, we conducted an ablation study by varying the buffer frequency and the minimum cluster size used in our model. In the baseline configuration, we used a 10 kHz buffer and a minimum cluster size of 5. For the study, we tested alternative values: minimum cluster sizes of 3, 5, and 7, and buffer widths of 5 kHz, 15 kHz, and 20 kHz. We then evaluated the impact of these variations on key image quality metrics, specifically FWHM and pSNR. All the ablations done with the drive frequency of 55kHz.

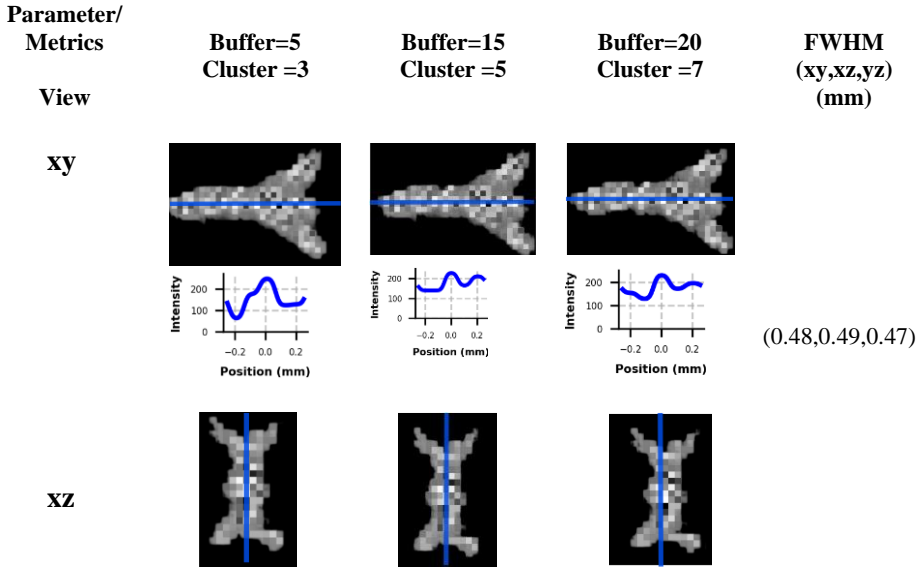

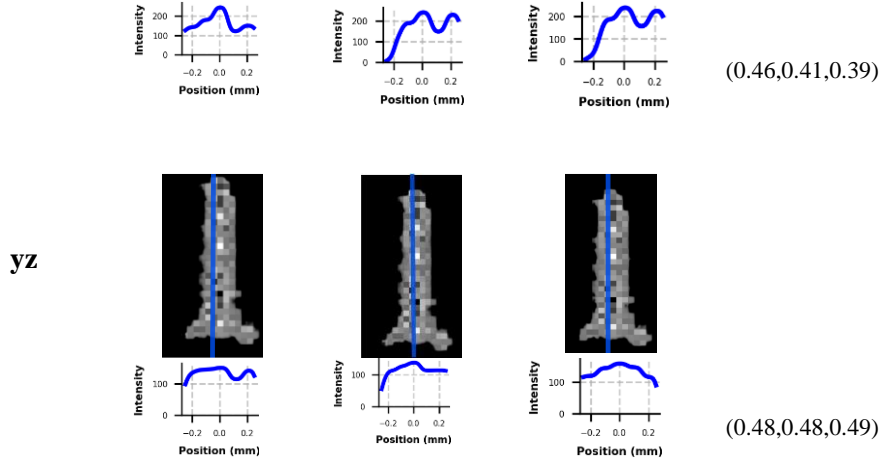

**Fig.15.** Reconstructed images at a drive frequency of 55 kHz using different buffer widths/cluster sizes (5 kHz/3, 15 kHz/5, and 20 kHz/7). The plots beneath each image show the corresponding FWHM profiles in the xy, xz, and yz planes.

**Table.10.** Quantitative comparison of reconstructed images under different buffer and cluster settings.

| Parameter | Buffer=5 Cluster =3 | | | Buffer=15 Cluster =5 | | | Buffer=20 Cluster =7 | | |
|---|---|---|---|---|---|---|---|---|---|
| Metrics View | pSNR(dB) | nRMSE | SSIM | pSNR(dB) | nRMSE | SSIM | pSNR(dB) | nRMSE | SSIM |
| xy | **46.85** | **0.0140** | **0.993** | 12.94 | 0.695 | 0.748 | 13.49 | 0.694 | 0.752 |
| xz | 13.04 | 0.7325 | 0.734 | 43.19 | 0.022 | 0.986 | 13.51 | 0.694 | 0.752 |
| yz | 13.98 | 0.6731 | 0.7575 | 16.39 | 0.509 | 0.821 | 43.95 | 0.021 | 0.988 |

The ablation study demonstrates how altering the buffer size and the minimum cluster size impacts the quality of the reconstructed images. Looking first at the xy view, as shown in Fig. 14, the smallest buffer (5 kHz, cluster size 3) clearly yields the best results. The images here are sharp and clean, supported by very high metric values (pSNR = 46.85 dB, SSIM = 0.993) as in Table 10. When the buffer is increased to 15 kHz or 20 kHz, the image quality significantly deteriorates, with pSNR values dropping to around 13 dB and SSIM also decreasing. Interestingly, the FWHM values (approximately 0.48–0.49 mm) remain almost unchanged, indicating that the resolution does not change significantly; however, the overall clarity and noise levels in the images deteriorate.

In contrast, the xz view behaves quite differently. Here, the mid-range buffer (15 kHz, cluster size 5) yields the clearest reconstruction, achieving a pSNR of 43.19 dB and an SSIM of 0.986. Both the smaller and larger buffer settings yield significantly lower image quality. The FWHM values (0.46, 0.41, 0.39 mm) suggest that the resolution actually improves slightly as the buffer increases, even though the noise behavior changes depending on the setting.

Finally, in the yz view, the highest buffer setting (20 kHz, cluster size 7) yields the best performance, achieving a strong pSNR of 43.95 dB and an SSIM of nearly 0.99. Smaller buffers yield weaker results, accompanied by increased noise and lower similarity to the ground truth. Unlike in the xz view, the FWHM values remain almost the same (around 0.48 mm), suggesting that resolution is stable in this orientation. Overall, these results indicate that the effect of buffer and cluster size varies across different views; small buffers are most effective for xy, while larger buffers are more effective in xz and yz.

### 3.4.2. Using even and mixed harmonics

To further investigate the influence of harmonic selection on image quality, we conducted a second ablation study focusing on the reconstructed images using only even harmonics and mixed (odd + even) harmonics. While the basic research focused on the role of odd harmonics, this study aims to assess how the selective inclusion of even harmonics, either alone or in combination with significant odd harmonics, affects the spatial resolution of the reconstructed images. To quantitatively evaluate the outcomes, we measured the FWHM of the central signal peaks in each image, providing a direct metric of image sharpness and resolution. The results indicate that the FWHM becomes larger compared to the initial study, where only odd harmonics were used for image reconstruction, as in Figure 16.

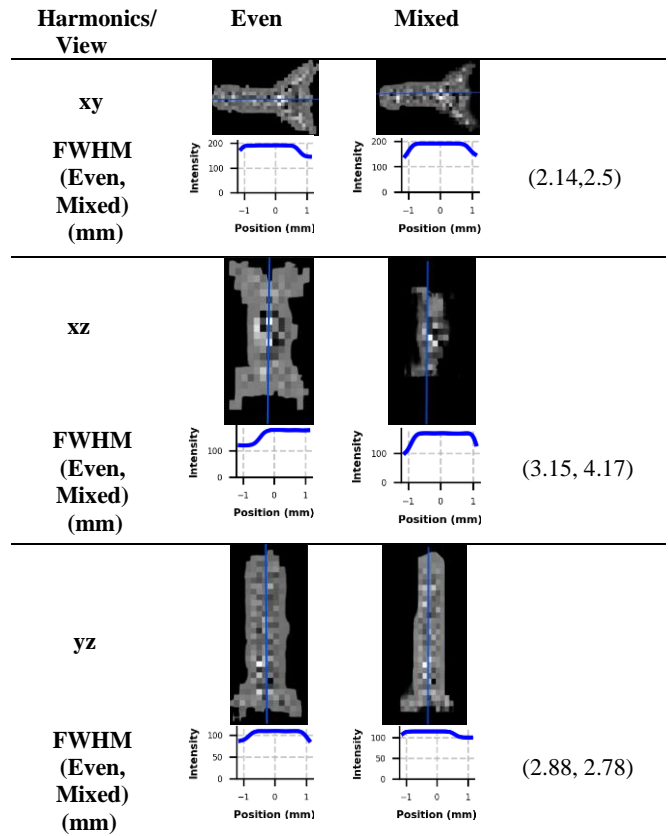

**Fig.16.** The reconstructed images at a drive frequency of 55 kHz using the signal reconstructed with even harmonics and mixed harmonics, the corresponding FWHM plots/values are also provided under each image.

## 4. Discussion

In MPI, image detail depends heavily on the strength and quality of the measured signals, but factors like nanoparticle type, coil design, noise, and reconstruction methods influence it. Despite progress, improving resolution still comes with challenges[13]. For example, magnetic relaxation effects can blur the signal, and designing the perfect coil often involves trade-offs between resolution and sensitivity.

Although gradient strength sets the baseline spatial resolution in MPI, our study shows that changing the drive frequency, even with a fixed amplitude, can meaningfully influence image quality. Higher frequencies enrich the induced voltage signal by introducing stronger and more numerous harmonics, which can enhance spatial detail. However, these frequencies also stress the nanoparticles' response limits, leading to relaxation effects that blur the image. These dynamics affect the PSF, making it sharper or broader depending on how well particles follow the fast-changing field. To isolate the influence of frequency alone and ensure consistent excitation strength across all tests, we fixed the drive field amplitude throughout the study. This allowed for a controlled analysis of how frequency-dependent magnetization dynamics shape spatial resolution.

Changing the drive frequency allows us to capture more harmonic signals from MNPs, which helps us better characterize their behavior. Since different types of MNPs respond in unique ways at a given frequency, using multiple frequencies makes these differences even clearer. When imaging several types of MNPs at once, this rapid frequency switching not only highlights their distinctions but also improves the overall spatial resolution of the images, giving us sharper and more detailed results[34]. This indicates that the drive frequency can influence the harmonics picked up by the receive coils, which in turn affects the spatial resolution of the images[35].

Operating at lower drive frequencies allows the MNPs more time to align with the changing magnetic field, producing a sharper PSF and enhancing spatial resolution, as in Fig. 6. However, the downside is that the reduced magnetization changes lead to weaker induced signals, potentially reducing the clarity of the reconstructed image[16]. These opposing trends emphasize the need for a frequency-optimized balance between signal strength and image resolution—one that our simulation framework is well-equipped to investigate.

By using CAHA to isolate and enhance harmonics that carry the most useful spatial information, we can recover high-resolution features that might otherwise be lost due to signal degradation at higher frequencies as seen in Fig.10. This harmonic-focused analysis allows for more accurate reconstructions even in challenging frequency conditions with low SNR as illustrated in Table 2, paving the way for improved MPI performance across a wider range of operating parameters. This is due to its ability to extract spatial information by examining how harmonics in each axis correspond to particle positions.

The frequency-domain reconstruction preserves the original phase $\varphi_k$ for each retained harmonic, which is crucial for reconstructing accurate spatial features. The use of a soft noise floor $\varepsilon$ in Equ. 12 for discarded components mitigates the risk of artifacts due to sharp frequency cutoffs. This results in cleaner, more stable images, particularly when comparing CAHA-enhanced reconstructions with those obtained from raw signals without harmonic selection.

In our implementation of CAHA, the original phase $\varphi_k$ of each harmonic is preserved during signal reconstruction. The adaptive weighting is applied exclusively to the harmonic magnitudes, ensuring that the temporal coherence of the reconstructed signal is maintained. While cross-axis weighting modifies the relative amplitude contributions of harmonics across x, y, and z, it does not introduce artificial phase shifts or inter-axis inconsistencies. Since MPI image reconstruction depends on the linear superposition of signals along each axis, retaining the true phase guarantees consistency in the encoded spatial information. Thus, CAHA enhances harmonic selectivity without compromising the phase integrity required for accurate reconstruction.

While higher frequencies generally improve resolution, as evidenced by reduced FWHM values, this benefit diminished or even reversed in some cases. At the highest drive frequency (85 kHz), a slight degradation in the FWHM for the reconstructed images R is observed in some views. This effect could stem from two interrelated phenomena: first, the increasing impact of nanoparticle relaxation effects at high frequencies applied in the model, which distort signal linearity and harmonic composition; second, the possibility that CAHA's noise suppression introduces minor smoothing artifacts, while it's helpful in low-SNR contexts, it may decrease sharp transitions in high-frequency ranges.

Although our simulations swept drive frequencies in the kHz range, while the Open MPI scanner runs at a fixed 2.5 MHz, the insights transfer because the same physical principles govern both regimes. In simple terms, increasing drive frequency always strengthens the signal and produces richer harmonics, but it also pushes nanoparticles closer to their relaxation limits, which can blur the image. By sweeping frequencies in simulation, we could clearly map out this trade-off and show how CAHA helps recover the most informative harmonics. When moving to the MHz regime of Open MPI, CAHA did not need to be redesigned; the adaptive thresholds automatically emphasized the harmonics that carried the most stable spatial information, mainly in the 200–300 kHz band. This shows that the method is flexible: it adapts to the operating frequency range, ensuring that the benefits seen in simulation remain valid in experimental systems as well.

The computational cost of our CAHA model (see Supplementary Material Section 2) remains reasonable, even with a large number of voxels in the simulated phantom (50,653). Despite the data size, key steps like signal simulation and image reconstruction were completed within practical time and memory limits. While some stages are more demanding, the overall workflow is efficient and well-suited for advanced MPI simulations.

Our study shows that optimizing drive frequency together with CAHA processing can sharpen MPI images, which in medicine could mean clearer views of tiny blood vessels, more precise tumor borders, or accurate tracking of drug delivery. This has the potential to support earlier diagnosis and more targeted treatments, ultimately improving patient outcomes. To bring these benefits into clinical use, however, several technical factors must be addressed: drive field settings need to balance image quality with patient safety limits, nanoparticle tracers must be carefully designed for both safety and responsiveness, and the imaging hardware and reconstruction algorithms must deliver reliable results quickly. By tackling these challenges, our findings move MPI closer to becoming a practical tool for patient care.

While the CAHA method shows clear improvements in spatial resolution and noise suppression, several limitations should be acknowledged. First, CAHA's performance depends on the accurate identification of high-SNR harmonics, which may vary across datasets and noise conditions. In cases of low particle concentration or weak signal strength, the clustering-based noise estimation may misclassify useful harmonics, potentially leading to over-smoothing or the loss of fine features. Additionally, CAHA assumes that the frequency content is sufficiently rich and spatially informative along all axes, which may not hold in systems with strong anisotropies or reduced sensitivity, particularly along the z-axis, as observed in our results. Moreover, the method's reliance on adaptive thresholds introduces sensitivity to parameter tuning, such as SNR percentile cutoffs or clustering settings, which may require calibration for different imaging setups. Lastly, while CAHA improves signal quality, it adds computational complexity through multi-axis spectral processing and dynamic harmonic weighting, which could pose challenges for real-time applications without further optimization.

## 5. Conclusion

This study presents a comprehensive MPI simulation framework that brings it closer to replicating realistic signal behavior and improving image reconstruction quality. At the heart of this framework is our signal processing technique, CAHA, which selectively extracts high-SNR harmonic components from the induced voltage signals. By targeting spatially informative harmonics across the x, y, and z axes, CAHA offers a fresh perspective on improving image resolution without compromising signal integrity.

**Data availability**

The data used in this work include simulated data generated using a custom code, which will be made available upon request, and the publicly available Open MPI data, accessible at: https://media.tuhh.de/ibi/openMPIData/data/calibrations.

**Acknowledgments**

This work was supported in part by the Key R&D Program of Shandong Province, China 2022CXGC010501 and the National Natural Science Foundation of China under Grant No. 82227802.

**References**

[1]  X. Han *et al.*, "The applications of magnetic particle imaging: From cell to body," Oct. 09, 2020, *Multidisciplinary Digital Publishing Institute (MDPI)*. doi: 10.3390/diagnostics10100800.

[2]  B. Rezaei *et al.*, "Magnetic nanoparticles for magnetic particle imaging (MPI): design and applications," May 24, 2024, *Royal Society of Chemistry*. doi: 10.1039/d4nr01195c.

[3]  N. Panagiotopoulos *et al.*, "Magnetic particle imaging: Current developments and future directions," *Int J Nanomedicine*, vol. 10, pp. 3097–3114, 2015, doi: 10.2147/IJN.S70488.

[4]  C. Billings, M. Langley, G. Warrington, F. Mashali, and J. A. Johnson, "Magnetic particle imaging: Current and future applications, magnetic nanoparticle synthesis methods and safety measures," Jul. 02, 2021, *MDPI*. doi: 10.3390/ijms22147651.

[5]  T. M. B. Knopp, Tobias, *Magnetic Particle Imaging:An introduction to imaging principles and scanner instrumentation*, no. July. 2020.

[6]  M. H. Pablico-Lansigan, S. F. Situ, and A. C. S. Samia, "Magnetic particle imaging: Advancements and perspectives for real-time in vivo monitoring and image-guided therapy," *Nanoscale*, vol. 5, no. 10, pp. 4040–4055, 2013, doi: 10.1039/c3nr00544e.

[7] B. Gleich and J. Weizenecker, "Tomographic imaging using the nonlinear response of magnetic particles," *Nature*, vol. 435, no. 7046, pp. 1214–1217, 2005, doi: 10.1038/nature03808.

[8] T. Knopp and A. Weber, "Sparse reconstruction of the magnetic particle imaging system matrix," *IEEE Trans Med Imaging*, vol. 32, no. 8, pp. 1473–1480, 2013, doi: 10.1109/TMI.2013.2258029.

[9] D. Roy, G. J. Steyer, M. Gargesha, M. E. Stone, and L. Wilson, "Exercise for Improving Age-Related Hyperkyphotic Posture: A Systematic Reveiw," vol. 292, no. 3, pp. 342–351, 2009, doi: 10.1016/j.jmmm.2009.02.083.Optimization.

[10] P. W. Goodwill and S. M. Conolly, "The X-space formulation of the magnetic particle imaging process: 1-D signal, resolution, bandwidth, SNR, SAR, and magnetostimulation," *IEEE Trans Med Imaging*, vol. 29, no. 11, pp. 1851–1859, 2010, doi: 10.1109/TMI.2010.2052284.

[11] C. Kuhlmann *et al.*, "Drive-Field Frequency Dependent MPI Performance of Single-Core Magnetite Nanoparticle Tracers," *IEEE Trans Magn*, vol. 51, no. 2, pp. 1–4, 2015, doi: 10.1109/TMAG.2014.2329772.

[12] J. Franke and J. Chacon-Caldera, "Magnetic particle imaging," *Magnetic Materials and Technologies for Medical Applications*, vol. 36, no. 7, pp. 339–393, 2021, doi: 10.1016/B978-0-12-822532-5.00015-7.

[13] J. Rahmer, J. Weizenecker, B. Gleich, and J. Borgert, "Signal encoding in magnetic particle imaging: Properties of the system function," *BMC Med Imaging*, vol. 9, pp. 1–21, 2009, doi: 10.1186/1471-2342-9-4.

[14] T. Q. Bui, M. A. Henn, W. L. Tew, M. A. Catterton, and S. I. Woods, "Harmonic dependence of thermal magnetic particle imaging," *Sci Rep*, vol. 13, no. 1, pp. 1–17, 2023, doi: 10.1038/s41598-023-42620-1.

[15] P. Vogel, M. A. Rückert, T. Kampf, and V. C. Behr, "Highly Flexible and Modular Simulation Framework for Magnetic Particle Imaging," vol. 5.

[16] Z. W. Tay, D. W. Hensley, P. Chandrasekharan, B. Zheng, and S. M. Conolly, "Optimization of Drive Parameters for Resolution, Sensitivity and Safety in Magnetic Particle Imaging," *IEEE Trans Med Imaging*, vol. 39, no. 5, pp. 1724–1734, 2020, doi: 10.1109/TMI.2019.2957041.

[17] L. R. Croft *et al.*, "Low drive field amplitude for improved image resolution in magnetic particle imaging," *Med Phys*, vol. 43, no. 1, pp. 424–435, 2016, doi: 10.1118/1.4938097.

[18] Z. W. Tay, P. W. Goodwill, D. W. Hensley, L. A. Taylor, B. Zheng, and S. M. Conolly, "A High-Throughput, Arbitrary-Waveform, MPI Spectrometer and Relaxometer for Comprehensive Magnetic Particle Optimization and Characterization," *Sci Rep*, vol. 6, no. March, pp. 1–12, 2016, doi: 10.1038/srep34180.

[19] Y. Shen *et al.*, "A systematic 3-D magnetic particle imaging simulation model for quantitative analysis of reconstruction image quality," *Comput Methods Programs Biomed*, vol. 252, no. May, p. 108250, 2024, doi: 10.1016/j.cmpb.2024.108250.

[20]  Y. Liu *et al.*, "Weighted sum of harmonic signals for direct imaging in magnetic particle imaging," *Phys Med Biol*, vol. 68, no. 1, 2023, doi: 10.1088/1361-6560/aca9b9.

[21]  Y. Liao *et al.*, "Improving the Resolution of Single-harmonic MPI Using Perpendicular Signal Transformation," *Int J Magn Part Imaging*, vol. 10, no. 1, pp. 1–4, 2024, doi: 10.18416/IJMPI.2024.2403030.

[22]  S. Shan *et al.*, "SPFS: SNR peak-based frequency selection method to alleviate resolution degradation in MPI real-time imaging," *Phys Med Biol*, vol. 69, no. 11, 2024, doi: 10.1088/1361-6560/ad3c90.

[23]  Y. Liu, G. Li, J. Li, Z. Tang, Y. An, and J. Tian, "Space-Specific Mixing Excitation for High-SNR Spatial Encoding in Magnetic Particle Imaging," *IEEE Trans Biomed Eng*, vol. 71, no. 10, pp. 2889–2899, 2024, doi: 10.1109/TBME.2024.3400274.

[24]  D. L. Donoho, "Compressed sensing," *IEEE Trans Inf Theory*, vol. 52, no. 4, pp. 1289–1306, 2006, doi: 10.1109/TIT.2006.871582.

[25]  T. Knopp, P. Szwargulski, F. Griese, and M. Gräser, "OpenMPIData: An initiative for freely accessible magnetic particle imaging data," *Data Brief*, vol. 28, Feb. 2020, doi: 10.1016/j.dib.2019.104971.

[26]  A. Beck and M. Teboulle, "A Fast Iterative Shrinkage-Thresholding Algorithm," *Society for Industrial and Applied Mathematics Journal on Imaging Sciences*, vol. 2, no. 1, pp. 183–202, 2009.

[27]  T. Knopp, K. Them, M. Kaul, and N. Gdaniec, "Joint reconstruction of non-overlapping magnetic particle imaging focus-field data," *Phys Med Biol*, vol. 60, no. 8, pp. L15–L21, 2015, doi: 10.1088/0031-9155/60/8/L15.

[28]  C. C. Paige and M. A. Saunders, "LSQR: An Algorithm for Sparse Linear Equations and Sparse Least Squares," *ACM Transactions on Mathematical Software (TOMS)*, vol. 8, no. 1, pp. 43–71, 1982, doi: 10.1145/355984.355989.

[29]  J. Lampe *et al.*, "Fast reconstruction in magnetic particle imaging," *Phys Med Biol*, vol. 57, no. 4, pp. 1113–1134, 2012, doi: 10.1088/0031-9155/57/4/1113.

[30]  M. Storath *et al.*, "Edge Preserving and Noise Reducing Reconstruction for Magnetic Particle Imaging," *IEEE Trans Med Imaging*, vol. 36, no. 1, pp. 74–85, 2017, doi: 10.1109/TMI.2016.2593954.

[31]  A. Beck and M. Teboulle, "Fast gradient-based algorithms for constrained total variation image denoising and deblurring problems," *IEEE Transactions on Image Processing*, vol. 18, no. 11, pp. 2419–2434, 2009, doi: 10.1109/TIP.2009.2028250.

[32]  N. Dey, "Springer Tracts in Nature-Inspired Computing Applied Multi-objective Optimization."

[33]  X. Chen *et al.*, "Simulation of reconstruction based on the system matrix for magnetic particle imaging," *Biomed Signal Process Control*, vol. 71, 2022, doi: 10.1016/j.bspc.2021.103171.

[34] H. Zhang *et al.*, "A Novel Hand-Held Multi-Color Magnetic Particle Imaging Device Based on Rapid Frequency Conversion," *IEEE Trans Biomed Eng*, vol. 71, no. 12, pp. 3602–3611, 2024, doi: 10.1109/TBME.2024.3434961.

[35] E. Mattingly *et al.*, "Design, construction and validation of a magnetic particle imaging (MPI) system for human brain imaging," *Phys Med Biol*, vol. 70, no. 1, Jan. 2025, doi: 10.1088/1361-6560/ad9db0.